\documentclass[twocolumn]{revtex4-1}

\pdfoutput=1

\usepackage{amsmath}
\usepackage{amsfonts}
\usepackage{amssymb}
\usepackage{graphicx}
\usepackage{graphics}
\usepackage{tikz}
\usetikzlibrary{shapes,snakes,decorations.pathreplacing,decorations.pathmorphing}
\definecolor{green}{cmyk}{0.5,0.0,0.6,0.3}

\begin{document}
\title{Fragmentation transitions in multi-state voter models}
\author{Gesa A. \surname{B\"ohme}}
\email{gesa@pks.mpg.de}
\affiliation{Max-Planck Institute for the Physics of Complex Systems, Dresden, Germany}
\author{Thilo \surname{Gross}}
\affiliation{University of Bristol, Department of Engineering Mathematics, Bristol, UK}
\date{\today}

\begin{abstract}
Adaptive models of opinion formation among humans can display a fragmentation transition, where a social network breaks into disconnected components. 
Here, we investigate this transition in a class of models with arbitrary number of opinions. 
In contrast to previous work we do not assume that opinions are equidistant or arranged on a one-dimensional conceptual axis. 
Our investigation reveals detailed analytical results on fragmentations in a three-opinion model, which are confirmed by agent-based simulations.
Furthermore, we show that in certain models the number of opinions can be reduced without affecting the fragmentation points.
\end{abstract}
\maketitle{}

\section{INTRODUCTION}
In many different fields networks have been used to describe and analyze complex systems consisting of interacting subunits. 
The applications of networks range from biological systems to technical devices and social communities \cite{Barabasi,NewmanRev,Bocca,dorogovnew}.
Accordingly, the building blocks of a network, the network nodes, can correspond to different entities, such as
 genes, neurons, computers, websites or individuals. The interactions among them, the links between the nodes, represent e.g. chemical reactions, physical connections, or social bonds. In the applications the temporal evolution of a network is often governed by two different types of dynamics: 
the dynamics \emph{on} the network, describing the evolution of the internal degrees of freedom, and the dynamics \emph{of} the network, capturing the evolution of the network topology.

Adaptive networks are characterized by an interplay of the dynamics on the network and the dynamics of the network, where neither of both types of dynamics can be neglected \cite{ThiloAdap,anbook}. 
It has been shown that this interplay gives rise to the emergence of complex topologies and dynamics \cite{zimmermann2000}, 
spontaneous appearance of different classes of nodes from an initially homogeneous population \cite{Kaneko, lydo}, and robust self-organization to critical states associated with phase transitions \cite{BornholdtRohlf,Meisel}. 
The self-organization of adaptive networks is believed to be of importance in the evolution of cooperation, opinion formation processes, epidemic dynamics, neural networks, and gene regulation \cite{wiki}. 

In the adaptive networks literature, opinion dynamics has currently attracted particular attention \cite{VazquezGeneralVM, HolmeNewman, KozmaBarrat, gesa, dirVM, guvenScience}. 
Typically, in these models a society is described as a network, where nodes correspond to individuals and links to social relationships. 
The internal state of the individuals indicate their position regarding some issue, such as political opinion, religious affiliation, or musical taste.
The individuals change their states by adopting opinions from their topological neighbors. The network topology, i.e.{} the specific pattern of nodes and links, changes as individuals break up relationships with dissenting neighbors and/or establish new relationships to those holding similar opinions.

The simplest adaptive-network model for opinion formation is the adaptive voter model \cite{VazquezGeneralVM, VazquezPcNodeup, GilZanetteAdaptVM, Nardini}. 
In this model and in many of its variants \cite{GilZanette, HolmeNewman, Kimura, guven} the relative rate of change of the topology compared to the change of node states is controlled by a single parameter, $p$, the \emph{rewiring rate}.
Depending on this parameter, the network either reaches a global homogeneous state, where
all nodes hold the same opinion or a fragmented state, consisting of two disconnected components which are internally homogeneous. The transition separating these two regimes is called a \emph{fragmentation transition}.

While most of the voter-like models only consider a binary choice of opinions, many real world situations offer a larger number of choices. 
In the physics literature some models for opinion formation, which consider arbitrary-many opinions have been studied \cite{HolmeNewman,Kimura}. In these models all opinions are ``equidistant'' in the sense that all interactions between any given pair of (different) 
opinions follow the same dynamical rules.
Models recognizing that the outcome of interactions may depend on a measure of similarity (or distance) between opinions are often considering an uncountable set of opinions and are therefore hard to treat analytically \cite{deffuant,KozmaBarrat}.
For instance in \cite{deffuant} opinions are placed on a 1-dimensional axis, resembling e.g.\,the political spectrum.

Here, we consider a natural extension of the original adaptive voter model, where we allow for an arbitrary countable set of opinions. In the  proposed model the rewiring rate that governs the interaction of conflicting agents is assumed to depend on the specific pairing of opinions held by the agents. The model can thus account for heterogeneous ``distances'' between opinions. 
A large distance, characterized by a high rewiring rate, indicates a controversial pairing, whereas a small distance, and correspondingly low rewiring rate, indicates that the respective opinions are almost in agreement.   

The proposed model is described in detail in Sec.~\ref{model}.
We then calculate the fragmentation diagram of a three-state model in Sec.~\ref{3states} and derive the corresponding fragmentation thresholds for systems with an arbitrary number of states in Sec.~\ref{gstates}. Finally, in Sec.~\ref{reduc}, we show that in certain systems a reduction is possible such that the dynamics can be captured by a system with a lower number of opinions.

\section{MULTI-STATE VOTER MODEL}
\label{model}
We consider a network of $N$ nodes, corresponding to individuals, and $L$ links, corresponding to social contacts.
Each node $\alpha$ holds a state $s_\alpha$, indicating the opinion held by the corresponding individual.
The network is initialized as a random graph with mean degree $\langle k \rangle=2L/N$. The initial node states are drawn randomly and with equal probability from the set of all states $\Gamma=\{g_1,g_2,g_3,\dots,g_G\}$, where the total number of states is
$|\Gamma|=G \ll N$.

The system is then updated as follows: In each update step a random link $(\alpha,\beta)$ is chosen. If $s_\alpha=s_\beta$ the link is said to be \emph{inert} and nothing happens. If $s_\alpha\neq s_\beta$, the link is said to be \emph{active} and an update occurs on the link.
A given update is either a rewiring event or an opinion adoption event, decided randomly depending on the similarity of the respective opinions.
For individuals $\alpha$, $\beta$ with opinions $s_\alpha=g_i$ and $s_\beta=g_j$, the update is an rewiring event with probability $p_{ij}$ and an opinion adoption event otherwise (probability $1-p_{ij}$). In the following the parameters $p_{ij}$ are called rewiring rates.
 
In a rewiring event the focal link $(\alpha,\beta)$ is severed, and a new link is created either from $\alpha$ to a randomly chosen node $\gamma$ with $s_\gamma=s_\alpha$, or from $\beta$ to a randomly chosen node $\gamma$ with $s_\gamma=s_\beta$. The choice between the two outcomes is made randomly with equal probability. In an opinion update, either node $\alpha$ changes its state to $s_\beta$ or node $\beta$ changes its state to $s_\alpha$, where the choice between both outcomes is again made randomly with equal probability. In the following we assume symmetric interactions, which implies $p_{ij}=p_{ji}$ such that the specific model is characterized by a set $\{p_{ij}\}$ of $G(G-1)/2$ parameters. 

The model proposed above preserves the symmetry of the direct interaction of two opinions postulated in the adaptive voter model, i.e.{}
in direct comparison no opinion is stronger then the other. However, it breaks the symmetry between different pairings of opinions such that rewiring is more likely in certain pairings than in others. 

From the adaptive voter model \cite{VazquezPcNodeup, VazquezGeneralVM} it is known that the fragmentation transition separates a so-called active regime from a fragmented regime. In the active regime a finite density of active links persists in the long-term dynamics such that there is ongoing dynamics until fluctuations drive the system eventually to an absorbing consensus state. In the fragmented regime, disconnected components emerge, which are internally in consensus.

Because the $G$-state model contains several different types of active links (corresponding to all possible pairings in $\Gamma$), regimes can occur where active links of a certain type vanish while others prevail. 
This can lead to configurations where a certain subset of the states only appears in one component of the network in which no state not belonging to this subset is present. In the following we call this situation a \emph{partially fragmented} state. 
In contrast to the \emph{fully fragmented} state where every component is internally in consensus, the dynamics in the partially fragmented state can continue in some components while others may be frozen in internal consensus. 
The partial fragmentation cannot be undone, so that achieving global consensus is impossible after partial fragmentation has occured. However, the ongoing dynamics in the active components will eventually lead to internal consensus in every component.
The absorbing state which is ultimately reached after a partial fragmentation therefore consists of $1<\gamma<G$ major components, holding the $\gamma$ surviving opinions, respectively. For $\gamma=G$ we recover full fragmentation and the case where
$\gamma=1$ we denote as the \emph{fully active} regime where all types of active links prevail. Only in the latter case (due to finite-size effects)  global consensus can be reached eventually.

\begin{figure}
 \begin{center}
\begin{tikzpicture}[scale=1]
\path[use as bounding box] (-1,2) rectangle (4.5,-2);

\draw[thick] (-1,2) rectangle (4.5,-2);

\tikzstyle{black}=[circle, thick,
                                    minimum size=1cm,
                                    draw=black!100,
                                    fill=black!100]

\tikzstyle{white}=[circle, thick,
                                    minimum size=1cm,
                                    draw=black!100,
                                    fill=white!1]

\tikzstyle{grey}=[circle, thick,
                                    minimum size=1cm,
                                    draw=black!100,
                                    fill=black!50]


\node[black](A) at (0,0){\Large{\textcolor{white}{A}}};
\node[white](B) at (3,1){\Large{B}};
\node[grey](C) at (3,-1){\Large{C}};

\path[-](A) edge[ultra thick, dashed] (B);
\path[-](A) edge[ultra thick, dashed] (C);
\path[-](C) edge[ultra thick] (B) ;

\draw (1.5,0.8) node {$p_1$};
\draw (1.5,-0.8) node {$p_2$};
\draw (3.3,0) node {$p_3$};

\draw[thick](3,0) ellipse (1.2cm and 1.7cm);

\end{tikzpicture}

\caption{State-network of a three-state voter model. The nodes in this network are the three opinions $A$, $B$ and $C$. 
The rewiring rates $p_1,p_2$ or $p_3$, encode the degree of controversy between agents holding the respective pairings of opinions $AB$, $BC$ or $CA$. 
The ellipse and dashed lines illustrate the example described in the text: 
 partial fragmentation with respect to $A$, i.e.{} the fragmentation of the corresponding networks of agents in a component where all individuals hold opinion $A$ and a second component where the individuals hold opinions $B$ and $C$. The ongoing dynamics in the latter component may later lead to the disappearance of either opinion $B$ or $C$.}
\label{states}
 \end{center}
\end{figure}
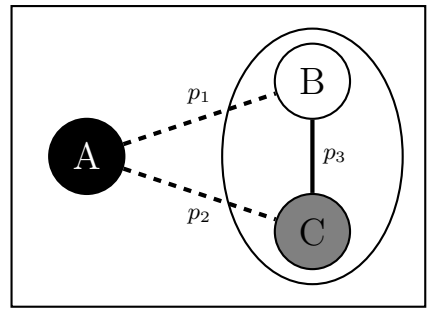

\section{FRAGMENTATION TRANSITIONS IN A THREE-STATE VOTER MODEL}
\label{3states}
We start our exploration of the proposed multi-state voter model by considering the case $G=3$, which is the simplest case which is not trivial ($G=1$) or extensively studied ($G=2$) \cite{VazquezPcNodeup, GuvenFederico, gesa}.
Let us consider the set of opinions $\Gamma=\{A,B,C\}$, giving rise to three different rewiring rates $\{p_{AB},p_{AC},p_{BC}\}$ which we denote as $\{p_1,p_2,p_3\}$ according to the state-network depicted in Fig.~\ref{states}.

Let us emphasize again the difference between the state-network and the network of individuals. The state-network is a complete, weighted graph with $G$ nodes which represents the relationships between different states, such that  
states connected by small rewiring rates are similar to each other, whereas states which are connected by large rewiring rates differ significantly from each other.
The network of individuals, in contrast, is an unweighted graph with $N$ nodes and mean-degree $\langle k \rangle$, which represents the interactions between individuals.

In principle, the three-state system can reach five different final states: fully active (leading eventually to global consensus), full fragmentation, and partial fragmentation with respect to $A$, $B$, or $C$.
Here, partial fragmentation with respect to a certain state refers to a situation where a component of nodes in that particular state fragments from an active component (a mixed component of nodes in the other states).

For the calculation of fragmentation thresholds we follow the approach given in \cite{gesa}. We determine the evolution equations for the number of \emph{active motifs} (network motifs containing active links) starting from a situation close to the fragmentation threshold. 
For simplicity, the present article uses only the simpler of two different active motif bases proposed in \cite{gesa}. We emphasize that all calculations below can also be carried out using more elaborate motif bases, but at the price of having to deal with considerably larger matrices. 

Following \cite{gesa}, we define $q$-fans as a bundle of $q$ active links of one type, say $AB$-links, connected to a single $A$- or $B$-node.
We do not account for the number of inert links connecting to this focal node.
For the sake of simplicity we also do not consider mixed active motifs containing all three states. 
We confirmed that effects of mixed motifs can be suitably captured by the procedure described below. 

\begin{figure}
 \begin{center}
 \includegraphics[trim=0.5cm 0cm 4cm 0cm,width=9cm]{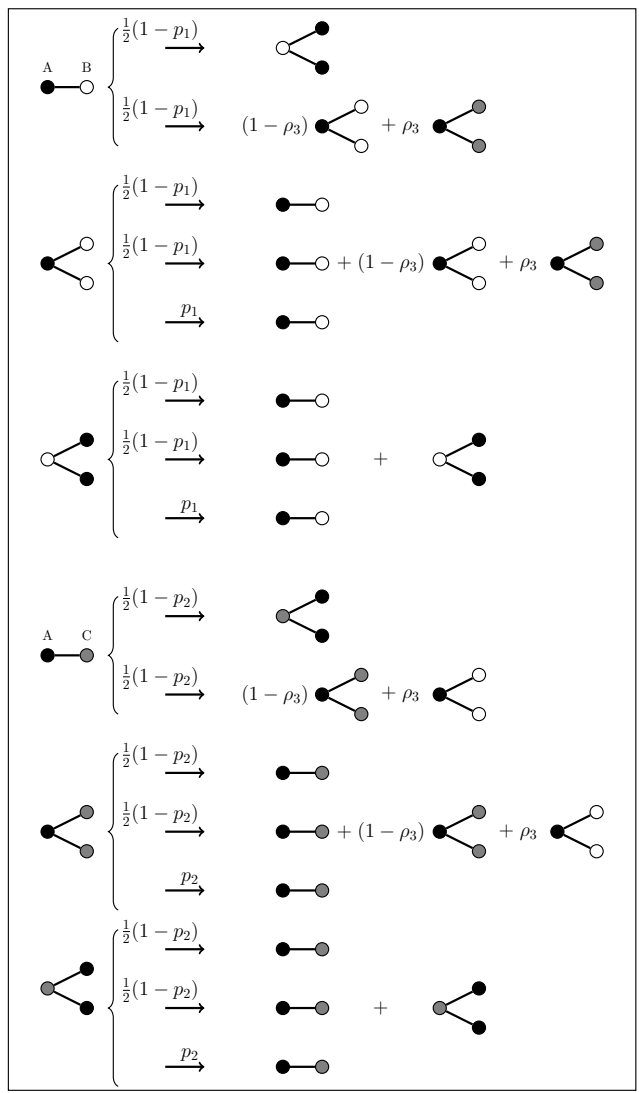}
\caption{Transitions of $AB$- and $AC$-fans for a degree-regular network with $k=3$ and equiprobable states for the scenario of partial fragmentation with respect to A. Black, white and grey nodes correspond to agents holding opinion $A$, $B$, and $C$, respectively.
The active link densities $\rho_1$ and $\rho_2$ are assumed to vanish close to the partial fragmentation point, whereas $\rho_3$, the densitiy 
of $BC$-links can be finite.}
\label{transitions}
 \end{center}
\end{figure}

We start by calculating the condition for partial fragmentation with respect to $A$ (see Fig.~\ref{states}).
The $A$-cluster fragments from the rest of the network when all $AB$-motifs and all $AC$-motifs vanish. 
Because in general $p_1\neq p_2$, we have to treat $AB$- and $AC$-motifs separately. 
We start by considering a network with two almost disconnected clusters, one of which is composed purely of $A$-nodes and the other of $B$- and $C$-nodes and then ask whether the fragmented state is stable, such that fragmentation is reached, or unstable, such that the system avoids fragmentation. 

In the almost fragmented state the expected effect of network updates on the active motifs is captured by a procedure proposed in \cite{gesa}. 
For the case of $k=3$ we obtain the transitions rules shown in Fig.~\ref{transitions}.
New active motifs are created when an opinion update occurs. We approximate the degree of the focal node $k$ by the networks mean degree $\langle k \rangle$.
Because of the clusters being almost-separated the newly formed active motif is a $k-1$-fan \cite{gesa}. This fan can subsequently lose active links due to subsequent opinion updates and rewiring events. 
We account for a finite density of active $BC$-links, $\rho_3$,  in the active component by creating an $AC$-fan ($AB$-fan) instead of an $AB$-fan ($AC$-fan) with probability $\rho_3$ when a new fan is created by a $B$-node ($C$-node) adopting opinion $A$.
 
If we start with equal distribution of states, the relation $\rho_3=[BC]/(k[B])=[BC]/(k[C])$ holds, where $[B]$ and $[C]$ denote the numbers of $B$-nodes and $C$-nodes, respectively. Note that $\rho_3$ differs from the global $BC$-link density $\rho_3^{(G)}=[BC]/L$.

The set of transitions for $k=3$ (see Fig.~\ref{transitions}) defines a dynamical system, describing the time evolution of the densities of active motifs close to the partial fragmentation with respect to $A$.
 The stability of the partially fragmented state in this system is then governed by the block-structured Jacobian,   
\begin{equation}
\label{jac3}
\mathbf{J}(p_1,p_2,\rho_{3})=
 \begin{pmatrix}
  \mathbf{D}_{p_1}-\mathbf{X}_{p_1}(\rho_{3})&\mathbf{X}_{p_1}(\rho_{3})\\
\mathbf{X}_{p_2}(\rho_{3})& \mathbf{D}_{p_2}-\mathbf{X}_{p_2}(\rho_{3})
 \end{pmatrix},
\end{equation}
where
\begin{equation}
\label{defd}
\mathbf{D}_p=
\begin{pmatrix}
 -1 & \tfrac{1}{2}(1-p) & \tfrac{1}{2}(1-p)\\
1 & \tfrac{1}{2}(1-p)-1 & 0\\
1  &  0  & \tfrac{1}{2}(1-p)-1
\end{pmatrix}
\end{equation}
and
\begin{equation}
\label{matx}
\mathbf{X}_p(\rho)=
\begin{pmatrix}
 0 & \tfrac{1}{2}(1-p)\rho & 0\\
 0 & \tfrac{1}{2}(1-p)\rho & 0\\
0 & 0 & 0
\end{pmatrix}.
\end{equation}

The diagonal blocks in the Jacobian given in Eq.\,(\ref{jac3}) can be interpreted as ``self-interaction'' terms, capturing contributions from the same motif-type, and the off-diagonal terms as ``exchange'' terms, capturing contributions from different motif-types. The structure of this Jacobian remains unchanged for any partial fragmentation of a three-state system, while the matrices (\ref{defd}) and (\ref{matx}) change when the motif set is altered. In particular, the dimension of these matrices increases with increasing mean degree and/or number of motifs considered.

\begin{figure}
 \begin{center}
 \includegraphics[width=.5\textwidth]{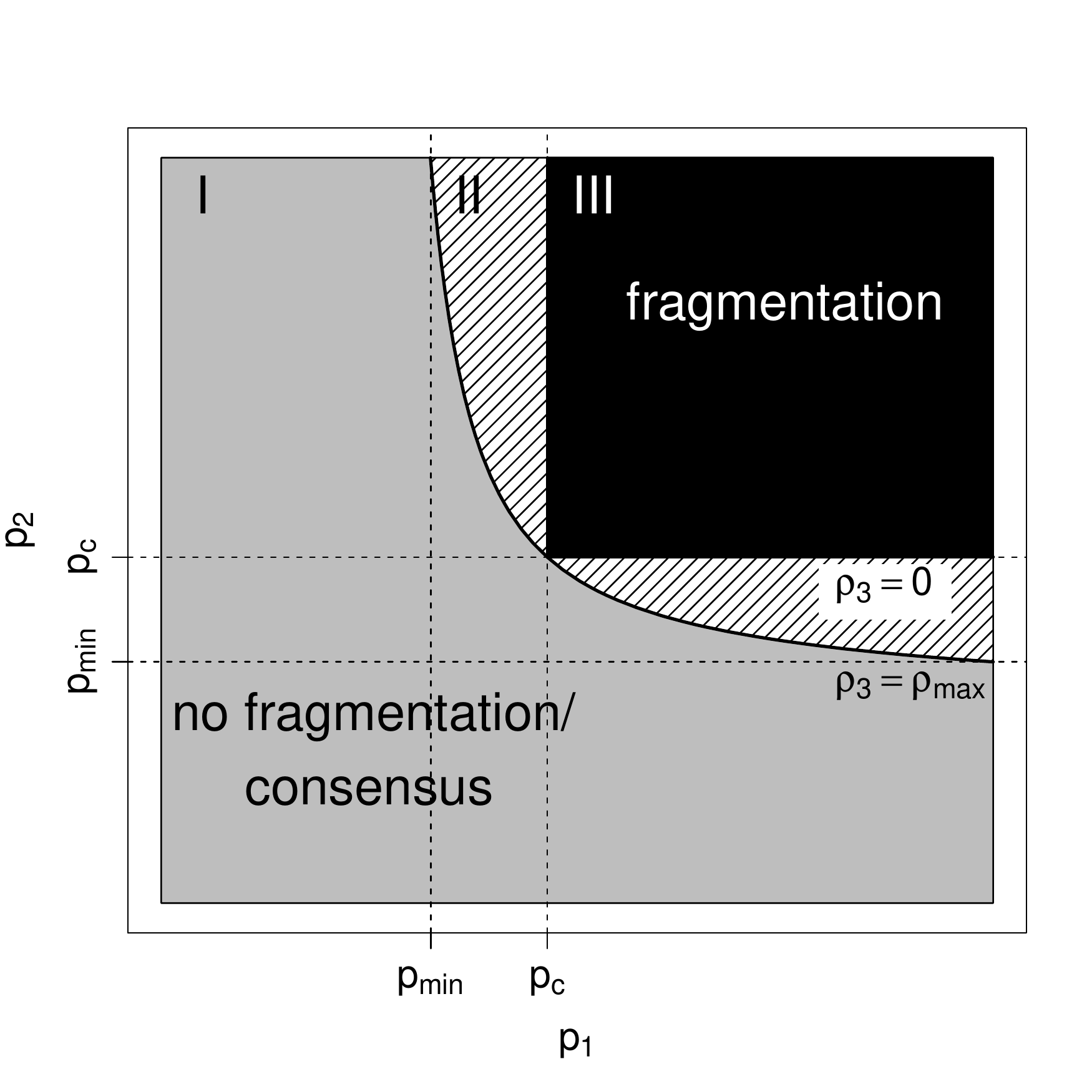}
\caption{Fragmentation diagram for a three-state system. 
Fragmentation with respect to opinion $A$ occurs always in the black region and never in the grey region. 
In the dashed region, fragmentation can occur depending on the value of $\rho_3$, the density of $BC$-links in the $BC$-component.  
If $p_1>p_3$ and $p_2>p_3$ then this fragmentation diagram characterizes the final state of the whole system. 
In this case region I coincides with the global consensus regime and full fragmentation is reached iff $p_3\ge p_c$,
where $p_c$ is the fragmentation threshold for the adaptive two-state voter model.}
\label{fragm_diag}
 \end{center}
\end{figure}

In a dynamical system a steady state is stable if all eigenvalues of the corresponding Jacobian have negative real parts \cite{kuznetsov1998}.
For the present system this means that the fragmented state is stable if all eigenvalues of the Jacobian $\mathbf{J}$ are negative and the fragmentation transition occurs as at least one of the eigenvalues acquires a positive real part. Therefore, 
demanding $\lambda(\mathbf{J})=0$, where $\lambda(\mathbf{J})$ is the leading eigenvalue of $\mathbf{J}$, yields a condition for the fragmentation transition, depending on the three parameters $p_1, p_2$, and $\rho_3$. 
The phase diagram in Fig.~\ref{fragm_diag} is a projection of this fragmentation condition on the $p_1$-$p_2$-plane for the extreme values of $\rho_3$. 

Let us first consider the case where $\rho_3=0$, which is encountered if $p_3$ exceeds $p_c$, the fragmentation threshold of the adaptive two-state voter model. 
In this case, $\mathbf{X}$ becomes zero and the set of eigenvalues of the Jacobian $\mathbf{J}$ is the conjunction of the eigenvalues of the matrices $\mathbf{D}_{p_1}$ and $\mathbf{D}_{p_2}$.
Thus, $\lambda(\mathbf{J})$ is negative iff $\lambda(\mathbf{D}_{p_1})$ and $\lambda(\mathbf{D}_{p_2})$ are negative.
Indeed matrices $\mathbf{D}_{p_1}$ and $\mathbf{D}_{p_2}$ are the Jacobians of the two uncoupled two-state systems $A-B$ and $A-C$.
Thus fragmentation of $A$ requires that the two-state fragmentation condition is met separately for the $AB$ and $AC$ subsystems. 
In other words, if the links between $B$ and $C$ nodes vanish ($\rho_3=0$), fragmentation occurs when both $p_1>p_c$ and $p_2>p_c$ (see Fig.~\ref{fragm_diag} region III).  

For studying the case $\rho_3>0$ we first note that every matrix-valued row of $\mathbf{J}$ sums to $\mathbf{D}_{p_i}$, where $i=1,2$.
Following \cite{deutsch}, as will be discussed below,  $\lambda(\mathbf{J})$ is bounded by $\lambda(\mathbf{D}_{p_1})$ and $\lambda(\mathbf{D}_{p_2})$. 
Therefore, fragmentation with respect to $A$ is guaranteed when $p_1 > p_c$ and $p_2 > p_c$, but can already occur when only 
either $p_1 > p_c$ or $p_2 > p_c$ is satisfied (see Fig.~\ref{fragm_diag} region II). 

The maximum extension of region II is observed when $p_3=0$, the corresponding maximal value of $\rho_3$, $\rho_{max}$, can be determined to good approximation by a moment closure approach (see Appendix), yielding
\begin{equation}
\label{rhoG3}
 \rho_{max}=\frac{k-1}{2k}.
\end{equation}
Solving the condition $\lambda(\mathbf{J}(p_1,p_2,\rho_{\max}))=0$ numerically yields the curve separating regions I and II in Fig.~\ref{fragm_diag}.
Moreover, from the diagram in Fig.~\ref{fragm_diag} it is clear that this curve implies the existence of a minimal rewiring rate $p_{\min}$, 
such that for $p_1<p_{\min}$ or $p_2<p_{\min}$ partial fragmentation with respect to $A$ becomes impossible.

Let us emphasize, that calculations of fragmentation thresholds for partial fragmentations build on the estimation of the active link densities from given rewiring rates. As there is no analytical expression for $\rho(p)$ in the whole $p$-range (\cite{GuvenFederico}) the long-term behavior can only be predicted with certainty in regions I and III of the fragmentation diagram.

\begin{figure}
\begin{center}
\begin{tabular}{cc}
\includegraphics[width=.24\textwidth]{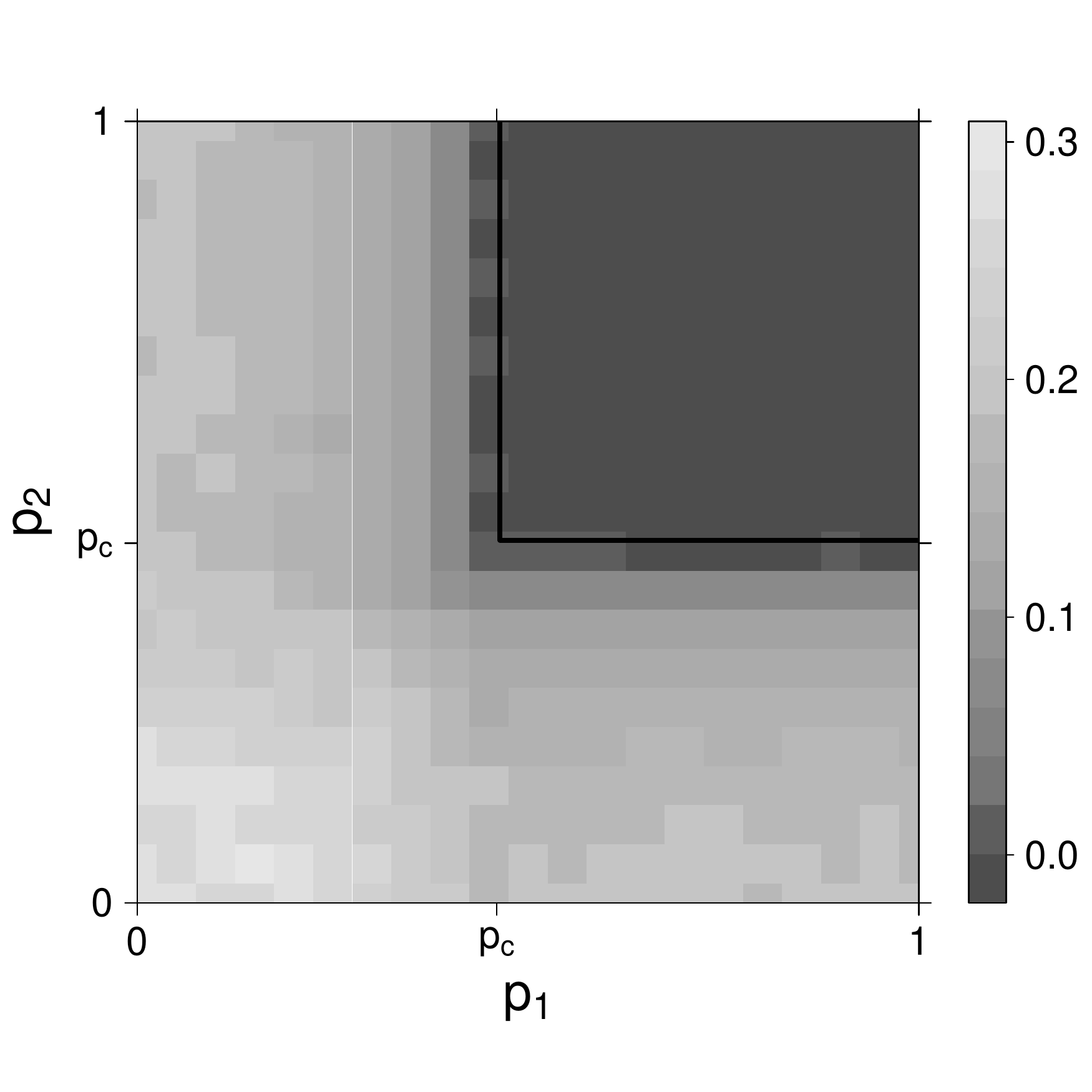}
&
\includegraphics[width=.24\textwidth]{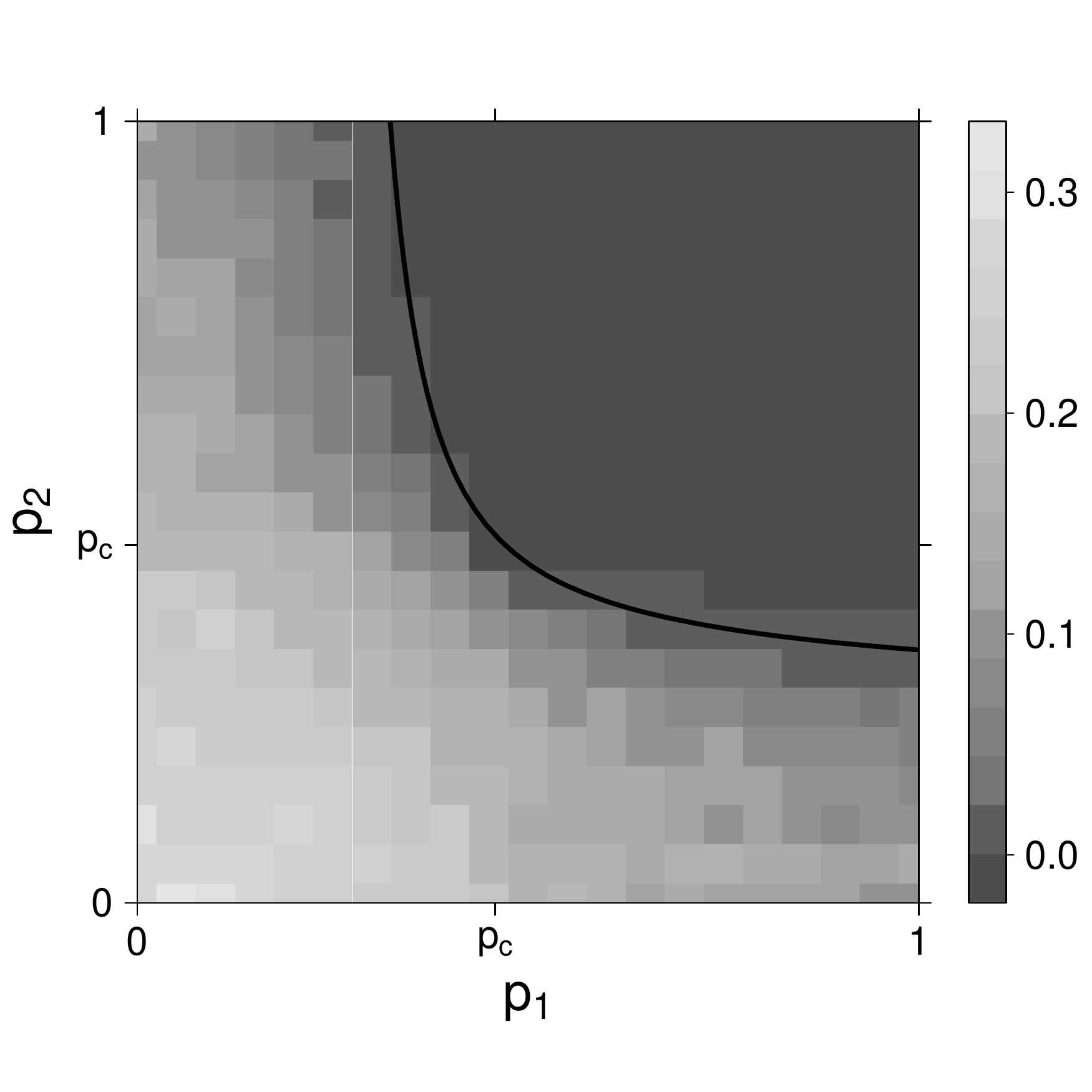}
\end{tabular}
\caption{Numerical phase diagram for the three-state model. Color-coded is the density of links, connecting the $A$ and the $BC$-cluster ($\rho_1+\rho_2$) over the rewiring rates $p_1$ and $p_2$. Dark grey regions correspond to fragmentation with respect to $A$.
The left panel shows the case $\rho_3=0\, (p_3=0.5)$. This corresponds to an uncoupled system: the critical rewiring rates for $p_1$ and $p_2$ are the same as for the two-state voter model, $p_c$.
In the right panel, $\rho_3$ is maximal ($p_3=0$). Here, the active link density in the active cluster leads to an extension of the fragmentation region.
 Black lines represent analytical results.
 $N$ = 10000, $\langle k \rangle$=4, averaged over 20 realizations.}
\label{simul}
\end{center}
\end{figure}

In summary, evaluating the partial fragmentation condition with respect to A, i.e. $\lambda(\mathbf{J})=0$, leads to a phase diagram as shown in Fig.~\ref{fragm_diag}, where three different regions can be distinguished:
In regions I and III partial fragmentation occurs or is avoided regardless of $p_3$, whereas in region II partial fragmentation depends on $\rho_3$ and consequently on the setting of the related rewiring rate $p_3$.
We found that these results are in very good agreement with data obtained from agent-based simulation of large networks (Fig.~\ref{simul}).

\begin{figure}
 \begin{center}
 \includegraphics[width=.5\textwidth]{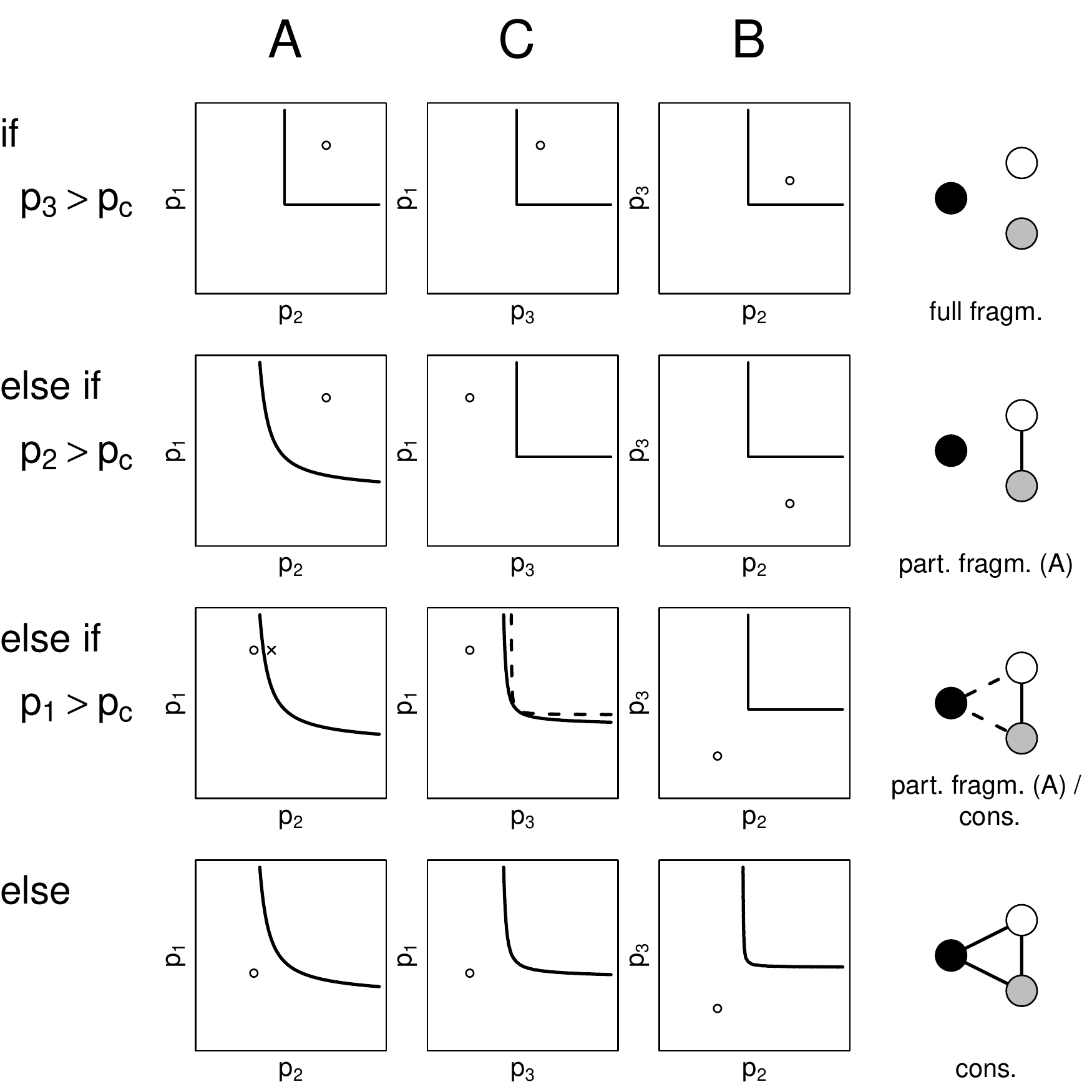}
\caption{Fragmentation diagrams for partial fragmentation with respect to state $A$, $C$ or $B$ for $p_1\ge p_2\ge p_3$. There are four different cases, according to the conditions given on the left-hand side of the chart.
The positions of the points $P=(p_2,p_1),Q=(p_3,p_1),R=(p_2,p_3)$ indicate for each case whether partial fragmentation is reached for the respective state. The final state of the whole system can be deduced from the outcomes for all three states and is
given on the right-hand side of the chart. For the first, second and forth case the final state can be predicted without ambiguity. In the third case, either partial fragmentation with respect to $A$ or consensus can be reached, depending on the
specific parameter setting (the two possibilities are indicated by two different symbols $\circ$ and $\times$ in the first diagram and a solid and a dashed line in the second diagram of the third row).
 It can be seen that partial fragmentation is only possible with respect to the state with the largest rewiring rates (here $A$).}
\label{fragm_tab}
 \end{center}
\end{figure}

Until now, we studied partial fragmentation with respect to one specific state (state $A$).
In order to predict the final state of the whole system, one has to analyze the corresponding partial fragmentation diagrams for each of the three states.  
Let us assume $p_1\ge p_2 \ge p_3$. Then, there are four cases to distinguish (see Fig.~\ref{fragm_tab}):
\begin{enumerate}
\item $p_3 > p_c$
\item $p_3 < p_c$ and $p_2>p_c$
\item $p_2<p_c$ and $p_1>p_c$
\item $p_1<p_c$.
\end{enumerate}
In case 1) full fragmentation is reached, because all points $P=(p_2,p_1)$, $Q=(p_3,p_1)$ and $R=(p_2,p_3)$ lie in the region III of their respective diagrams. 
In case 2) $P$ lies in III and $Q$ and $R$ in I. Thus, partial fragmentation with respect to $A$ occurs, while $B$- and $C$-nodes form an active cluster and in a finite system eventually reach consensus.
In case 3) the point $P$ lies either in region I or II of the corresponding fragmentation diagram for $A$, whereas $Q$ is always in region I, because $p_2\ge p_3$ and $\rho_3\ge \rho_2$.
This means that in this case either partial fragmentation with respect to $A$ or consensus is reached depending on the specific values of $p_1,p_2,p_3$. 
In case 4) $P$, $Q$ and $R$ lie in the region I of their respective diagrams and global consensus is reached.
Note that this shows that partial fragmentation can only occur with respect to that state, which is connected via the largest rewiring rates to the two other states in the state network.

In summary, we showed that in the three-state voter model either consensus, partial fragmentation or full fragmentation occurs. 
Full fragmentation is only reached when all rewiring rates exceed $p_c$. Analyzing the phase diagram with respect to the state which is connected by the largest rewiring rates to the other states suffices for the prediction of the final state of the whole system.
For quantitative predictions in region II of the diagram the active link density corresponding to the lowest rewiring rate has to be known. Qualitatively, one can say that if partial fragmentation occurs, then with respect to the ``most different'' state.

\section{FRAGMENTATION TRANSITIONS IN A $G$-STATE VOTER MODEL}
\label{gstates}

\begin{figure}
 \begin{tikzpicture}[scale=0.95]

\path[use as bounding box] (-1.5,2.5) rectangle (7.5,-3.5);

\draw[thick] (-1.5,2.5) rectangle (7.5,-3.5);

\tikzstyle{grey}=[circle, thick,
                                    minimum size=0.2cm,
                                    draw=black!100,
                                    fill=black!50]



\node[grey](A) at (-0.4,0.4){};
\node[grey](B) at (0,-2){};
\node[grey](C) at (1.8,0){};
\node[grey](D) at (1,-2){};
\node[grey](E) at (0.2,1.5){};
\node[grey](F) at (0.7,-0.8){};
\node[grey](G) at (-0.5,-0.9){};

\path[-](A) edge[thick] (B);
\path[-](A) edge[thick] (C);
\path[-](A) edge[thick] (D);
\path[-](A) edge[thick] (E);
\path[-](A) edge[thick] (F);
\path[-](A) edge[thick] (G);
\path[-](B) edge[thick] (C);
\path[-](B) edge[thick] (D);
\path[-](B) edge[thick] (E);
\path[-](B) edge[thick] (F);
\path[-](B) edge[thick] (G);
\path[-](C) edge[thick] (D);
\path[-](C) edge[thick] (E);
\path[-](C) edge[thick] (F);
\path[-](C) edge[thick] (G);
\path[-](E) edge[thick] (F);
\path[-](E) edge[thick] (G);
\path[-](F) edge[thick] (G);

\node[grey](B) at (-0.5+5,1){};
\node[grey](C) at (1.8+5,0){};
\node[grey](D) at (1+5,-1.8){};
\node[grey](E) at (0.2+5,1.5){};
\node[grey](G) at (-0.2+5,-0.8){};

\path[-](B) edge[thick] (C);
\path[-](B) edge[thick] (D);
\path[-](B) edge[thick] (E);
\path[-](B) edge[thick] (G);
\path[-](C) edge[thick] (D);
\path[-](C) edge[thick] (E);
\path[-](C) edge[thick] (G);
\path[-](E) edge[thick] (G);

\draw[font=\large] (1.1,1.3) node {$\{\rho_{ij}\}$};
\draw[font=\large] (6.2,1.2) node {$\{\rho_{\underline{ij}}\}$};
\draw[font=\large] (3,1) node {$\{p_{i\underline{i}}\}$};

\draw[-, ultra thick,dashed](2.3,0)--(3.9,0.1);
\draw[-, ultra thick,dashed](2.2,-0.5)--(3.8,-0.4);
\draw[-, ultra thick,dashed](2.2,0.2)--(3.9,0.8);
\draw[-, ultra thick,dashed](2.1,0.7)--(3.8,0.5);
\draw[-, ultra thick,dashed](2.1,-1.4)--(3.9,-.7);
\draw[-, ultra thick,dashed](2.2,-0.1)--(3.9,-1);
\draw[-, ultra thick,dashed](2.2,-0.85)--(4,-1.2);

\draw[thick](0.6,-0.3) ellipse (1.6cm and 2.3cm);
\draw[thick](5.6,-0.05) ellipse (1.7cm and 2.2cm);

\draw[font=\large] (6,-3) node {$G-s$ states};
\draw[font=\large] (0.5,-3) node {$s$ states};

\end{tikzpicture}
\caption{schematic representation of a state-network with $G$ states. Ellipses illustrate a fragmentation into two (possibly) active clusters of $s$ and $G-s$ states.
Dashed lines correspond to inter-cluster links with rewiring rates $\{p_{i\underline{i}}\}$, connecting every state in one cluster with every state in the other cluster.
The number of inter-cluster links, $s(G-s)$, determines the dimension of the (matrix-valued) Jacobian. 
Solid lines correspond to intra-cluster links within both clusters with active link densities $\{\rho_{ij}\}$ and $\{\rho_{\underline{ij}}\}$, respectively.
}
\label{gen2comp}
\end{figure}
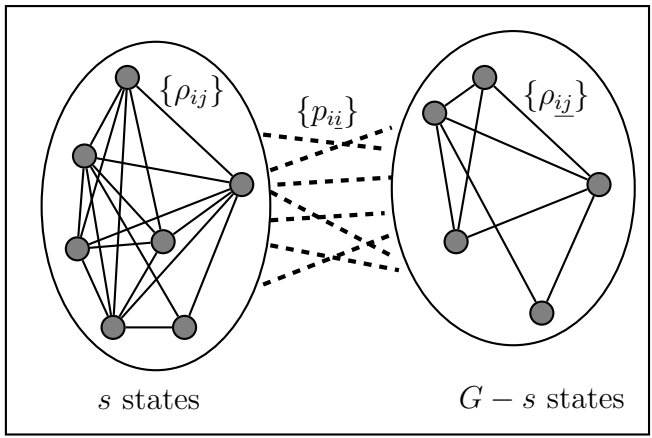

Let us now consider a general system of $G$ states. 
In contrast to the previous system partial fragmentations can also occur with respect to a group of states. A general multi-state network can thus fragment into several active components. 
Let us therefore calculate the condition for a system to fragment into two components containing $s$ and $G-s$ states, respectively (see Fig.~\ref{gen2comp}). This is in principle no restriction, as a fragmentation into more than two components can be treated as a 
fragmentation into two components where the active components in their turn fragment. 

For clarity we only use one level of indices from now on: we write $\mathbf{D}_{ij}$ and $\mathbf{X}_{ij}$ instead of $\mathbf{D}_{p_{ij}}$ and $\mathbf{X}_{p_{ij}}$.
Furthermore, in order to distinguish indices which refer to one component from those refering to the other component we use indices $i\in\{1,\dots,s\}$ for the component with $s$ states and 
indices $\underline{i}\in\{\underline{1},\dots,\underline{s}\}$ for the component with $\underline{s}=G-s$ states.
For example, the inter-component rewiring rates are then denoted as $p_{i\underline{i}}$ and intra-component active link densities as $\rho_{ij}$ and $\rho_{\underline{ij}}$, respectively. 

In analogy to the treatment of the three-state model, we consider a situation where the two 
clusters are almost fragmented. We then determine the evolution equations for a set of active motifs connecting the two components. In the three-state case these were of two types, $AB$- and $AC$-fans, which
led to a Jacobian of $2\times 2$ matrix-valued entries and a fragmentation condition which was a function of the rewiring rates $p_1$ and $p_2$ and the active link density $\rho_3$.
In the general case the Jacobian contains $s\underline s\times s \underline s$ matrix-valued entries, according to the number of inter-component links in the state-network (see Fig.~\ref{gen2comp}) 
and the fragmentation condition is a function of all inter-component rewiring rates $\{p_{i \underline i}\}$ and
all intra-component active link densities $\{\rho_{ij}\}$ and $\{\rho_{\underline{ij}}\}$.

Following the same procedure as for the three-state system, one finds that the general Jacobian exhibits a block-structure of $s \times s$ submatrices,
\begin{align}
\label{genjac}
&\mathbf{J}\bigl(p_{i\underline{i}},\rho_{ij},\rho_{\underline{ij}}\bigr)=\\
&\small{
 \begin{pmatrix}
\mathbf{\Delta}_{1}\bigl(\rho_{\underline{ij}},\rho_{1j}\bigr)& \mathbf{\xi}_{1}\bigl(\rho_{12}\bigr) & \cdots & \mathbf{\xi}_{1}\bigl(\rho_{1s}\bigr) \\
\mathbf{\xi}_{2}\bigl(\rho_{21}\bigr) & \mathbf{\Delta}_{2}\bigl(\rho_{\underline{ij}},\rho_{2j}\bigr)& \ddots & \vdots  \\
\vdots &\ddots&\ddots & \mathbf{\xi}_{(s-1)}\bigl(\rho_{(s-1)s}\bigr)\\
\mathbf{\xi}_{s}\bigl(\rho_{s1}\bigr) & \cdots & \mathbf{\xi}_{s}\bigl(\rho_{s(s-1)}\bigr)&\mathbf{\Delta}_{s}\bigl(\rho_{\underline{ij}},\rho_{sj}\bigr) 
\end{pmatrix}
}
,\nonumber
\end{align}
where $\mathbf{\Delta}_i$ and $\mathbf{\xi}_i$ are matrices of \mbox{$\underline s \times \underline s$} matrix-valued entries,
\begin{align*}
&\mathbf{\Delta}_{i}\bigl(\rho_{\underline{ij}},\rho_{ij}\bigr)=\\
&\small{
\begin{pmatrix}
\mathbf{\hat{D}}_{i\underline{1}}\bigl(\rho_{\underline{1j}},\rho_{ij}\bigr)& \mathbf{X}_{i\underline{1}}\bigl(\rho_{\underline{12}}\bigr) &\cdots & \mathbf{X}_{i\underline{1}}\bigl(\rho_{\underline{1s}}\bigr) \\
\mathbf{X}_{i\underline{2}}\bigl(\rho_{\underline{21}}\bigr) & \mathbf{\hat{D}}_{i\underline{2}}\bigl(\rho_{\underline{2j}},\rho_{ij}\bigr)& \ddots & \vdots  \\
\vdots &\ddots&\ddots & \mathbf{X}_{i\underline{s-1}}(\rho_{\underline{(s-1)s}})  \\
\mathbf{X}_{i\underline{s}}\bigl(\rho_{\underline{s1}}\bigr) & \cdots & \mathbf{X}_{i\underline{s-1}}(\rho_{\underline{s(s-1)}}) &\mathbf{\hat{D}}_{i\underline{s}}\bigl(\rho_{\underline{sj}},\rho_{ij}\bigr) 
\end{pmatrix}}
\end{align*}
and
\begin{align*}
 \mathbf{\xi}_{i}\bigl(\rho_{ij}\bigr)=
\small{
\begin{pmatrix}
\mathbf{X}^\prime_{i\underline{1}}\bigl(\rho_{ij}\bigr)&\mathbf{0}&\cdots & \mathbf{0}\\
\mathbf{0}&\mathbf{X}^\prime_{i\underline{2}}\bigl(\rho_{ij}\bigr)&\ddots&\vdots \\
\vdots &\ddots&\ddots & \mathbf{0} \\
\mathbf{0}&\cdots &\mathbf{0}& \mathbf{X}^\prime_{i\underline{s}}\bigl(\rho_{ij}\bigr) \\
\end{pmatrix}.
}
\end{align*}
Here, we introduced the abbreviation
\begin{equation*}
\mathbf{\hat{D}}_{i\underline{i}}\bigl(\rho_{\underline{ij}},\rho_{ij}\bigr)=
\mathbf{D}_{i\underline{i}}-\sum_{\underline{j}=\underline{1},\underline{j}\neq \underline{i}}^{\underline{s}}\mathbf{X}_{i\underline{i}}(\rho_{\underline{ij}})
-\sum_{j=1,j\neq i}^{s}\mathbf{X^\prime}_{i\underline{i}}(\rho_{ij}).
\end{equation*}
The matrices $\mathbf{D}$ and $\mathbf{X}$ for $k=3$ were already given in (\ref{defd}) and (\ref{matx})
and the matrix $\mathbf{X}^\prime$ for $k=3$ is,
\begin{equation}
\label{matxi}
 \mathbf{X^\prime}_{i\underline{i}}(\rho_{ij})=\begin{pmatrix}
 0 & 0& \tfrac{1}{2}(1-p_{i\underline{i}})\rho_{ij} \\
 0 & 0& 0\\
0 & 0 & \tfrac{1}{2}(1-p_{i\underline{i}})\rho_{ij}
\end{pmatrix}.
\end{equation}
The latter matrix appears for fragmentations where both of the fragmenting components are active, i.e.{} for $1<s<G-1$.

Note that every matrix-valued row of the Jacobian sums to $\mathbf{D}_{i\underline{i}}$ and refers to one specific type of inter-component link with rewiring rate $p_{i\underline{i}}$ in the state-network.
One such row thus represents the transitions for one motif-type. 
Entries $\mathbf{\hat{D}}$ on the diagonal capture the creation of motifs of the same type, while off-diagonal entries $\mathbf{X}$ and $\mathbf{X}^\prime$ denote transitions to different motif-types, which arise from the intra-component
link densities $\rho_{\underline{ij}}$ and $\rho_{ij}$, respectively. 
For example, the entries in one row describing the transitions of $g_1g_{\underline{2}}$-fans depend on the rewiring rate between the states $g_1$
and $g_{\underline 2}$, which is $p_{1\underline{2}}$, the active link densities between $g_1$ and all other states in the first component, $\rho_{1x}$, with $x\in\{2,\dots,s\}$; and the active link densities between $g_2$ and all remaining states in 
the second component, $\rho_{\underline{2y}}$, with $y\in \{\underline{1},\underline{3},\dots,\underline{s}\}$. 

In analogy to the three-state model the active link densities $\rho_{ij}$ entering in the Jacobian relate to the global active link density $\rho_{ij}^{(G)}$ as
\begin{equation}
\label{rhodef}
 \rho_{ij}=\frac{[g_ig_j]}{k[g_i]}=\frac{G}{2}\frac{[g_ig_j]}{L}=\frac{G}{2}\rho_{ij}^{(G)}.
\end{equation}
This holds analogously for $\rho_{\underline{ij}}$.

Stability analysis of the general Jacobian in (\ref{genjac}) is in principle possible, but leads to a fragmentation condition which depends directly or indirectly (through the active link densities) on all $G(G-1)/2$ different rewiring rates.
In contrast to the estimation of the active link density in a two-state system, in a multi-state system the active link density of a certain link-type does not only depend on the rewiring rate of that specific link-type, but also on the rewiring rates and active link densities of the neighboring links in the state-network.
Inferring the link densities analytically from the rewiring rates is presently an unsolved challenge.
So, even for given rewiring rates it is in general not possible to make quantitative predictions about fragmentation thresholds. Nevertheless, the structure of the Jacobian allows for qualitative predictions, which will be shown in the next section.

\section{REDUCTION PRINCIPLES FOR SPECIAL STATE-NETWORK TOPOLOGIES}
\label{reduc}
In this section we use theorems about upper and lower bounds of the largest eigenvalue $\lambda(\mathbf{M})$ of a nonnegative irreducible matrix $\mathbf{M}$.
The well-known Frobenius inequality states
\[
 \min_k S_k \le \lambda(\mathbf{M})\le \max_k S_k,
\]
where $S_i$ is the rowsum of the $i$-th row of $\mathbf{M}$. 
A generalization of the above inequality for a partitioned nonnegative irreducible square matrix $\mathbf{M}$ is given in \cite{deutsch}. Let us assume that $\mathbf{M}$ can be partitioned into square submatrices $\mathbf{M}_{ij}$, such that
\[
 \mathbf{M}=
\begin{pmatrix}
 \mathbf{M}_{11} & \mathbf{M}_{12} & \cdots & \mathbf{M}_{1N}\\
\mathbf{M}_{21} & \mathbf{M}_{22} & \cdots & \mathbf{M}_{2N}\\
\vdots & \vdots & \ddots & \vdots \\
\mathbf{M}_{N1} & \mathbf{M}_{N2} & \cdots & \mathbf{M}_{NN}
\end{pmatrix}.
\]
We define
\[
 \mathbf{S}_k=\sum_{j=1}^{N}\mathbf{M}_{kj}, \qquad k=1,\dots N
\]
as generalized, matrix-valued rowsums of $\mathbf{M}$. Then, the following inequality holds \cite{deutsch}:
\begin{equation}
\label{ineq}
\lambda(\min_k \mathbf{S}_k) \le \lambda(\mathbf{M})\le \lambda(\max_k \mathbf{S}_k).
\end{equation}
The expressions $\min_k$ and $\max_k$ have to be understood element-wise, i.e. the matrix $\min_k \mathbf{S}_k$ is the matrix which is obtained when we take element-wise the minimum over all $\mathbf{S}_k$ and
analogously for the maximum.

In the following we apply the theorem quoted above to the Jacobian given in \eqref{genjac}.
This is possible because $\mathbf{J}$ can be written as $\mathbf{J}=\mathbf{T}-\mathbf{1}$, where $\mathbf{T}$ is a nonnegative irreducible matrix and $\mathbf{1}$ is the identity matrix of appropriate dimension. 
We will consider two different partitions.

First, let us consider a partition (P1) of the Jacobian into $s\underline s$ submatrices. Then, the matrix-valued rowsums corresponding to this partition yield
\begin{align*}
\mathbf{S}_k^{(\rm{P1})}&=\mathbf{\hat{D}}_{i\underline{i}}\bigl(\rho_{\underline{ij}},\rho_{ij}\bigr)+\sum_{\underline{j}=\underline{1},\underline{j}\neq \underline{i}}^{\underline{s}}\mathbf{X}_{i\underline{i}}(\rho_{\underline{ij}})
+\sum_{j=1,j\neq i}^{s}\mathbf{X^\prime}_{i\underline{i}}(\rho_{ij}) \\
&= \mathbf{D}_{i\underline{i}},\qquad k=1,\dots,s\underline{s}.
\end{align*}
The matrices $\mathbf{D}_{i\underline{i}}$ only depend on $p_{i\underline i}$ and
it can be seen from \eqref{defd} that all non-constant entries in $\mathbf{D}_p$ increase with decreasing $p$. Therefore, we get for the upper and lower bounds of $\lambda(\mathbf{J})$
\begin{equation}
\label{ineqjac}
 \lambda(\mathbf{D}_{p_{\max}}) \le \lambda(\mathbf{J})\le \lambda(\mathbf{D}_{p_{\min}}),
\end{equation}
where
\[
 p_{\max}=\max_{i,\underline i} p_{i\underline i}, \qquad  p_{\min}=\min_{i,\underline i} p_{i\underline i}.
\]
From (\ref{ineqjac}) we deduce the following statements:
\begin{itemize}
 \item When all inter-cluster rewiring rates $p_{i\underline i}$ are below the threshold $p_c$, no fragmentation occurs, because $ \lambda(\mathbf{D}_{p_{\max}}) > 0$.
\item When all inter-cluster rewiring rates $p_{i\underline i}$ exceed the threshold $p_c$, fragmentation occurs, because $ \lambda(\mathbf{D}_{p_{\min}}) <0$.
\item If $p_{\min}= p_{\max}$, necessarily all inter-cluster rewiring rates must be equal. In that case, the fragmentation condition is the classical condition of the two-state voter model, $\lambda(\mathbf{J})=\lambda(\mathbf{D}_p)=0$,
which yields the critical rewiring rate $p_c$.
\end{itemize}
The first two results represent an intuitive generalization of our findings for the three-state case. The last result implies that if all inter-cluster rewiring rates are equal then the value of these rewiring rates, $p$, is the only parameter parameter on 
which the fragmentation transition depends. 
In this case a precise analytical estimation of the fragmentation point is possible because the active link densities arising from the intra-cluster links, do not enter. Furthermore note that this result is independent of the number of opinions. This implies that in the special case of equal inter-cluster rewiring rates systems of any size behave identically to a properly-initialized adaptive two-state voter model.  

\begin{figure}
\begin{center}
\input{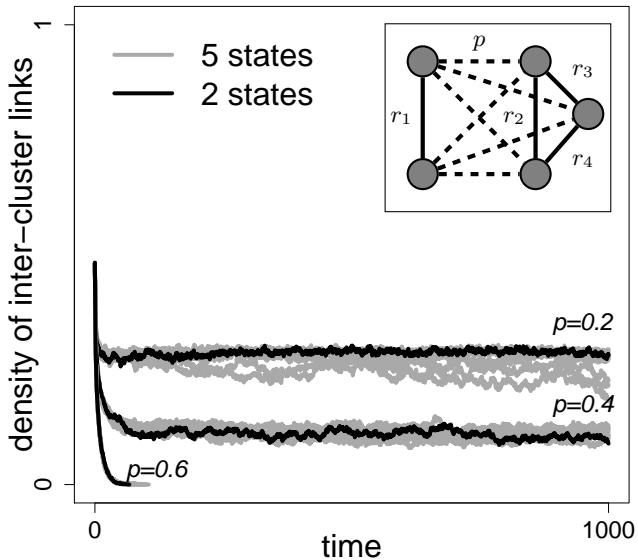}
\end{center}
\caption{Numerical test of the reduction principle. We run simulations for a 5-state system assigning random values to the rewiring rates $r_i$ (inset). Plotted is the density of inter-cluster links (dashed links) for three different values of $p$. 
For each $p$-value we run ten independent simulations with randomly chosen $r_i$ and compare the corresponding inter-cluster link density to the active link density of a two-state model with the same rewiring rate and a ratio 2:3 for the number of nodes in
opposite states. It can be seen that the inter-cluster density does not depend on $r_i$ and that the steady-state value equals the corresponding inter-cluster density (active link density) in the original two-state voter model. $N=10000,\langle k \rangle =4$.}
\label{5states}
\end{figure}

We test the latter result in a five-state system, considering a fragmentation into two components of 2 and 3 states, respectively (inset in Fig.~\ref{5states}).
 Simulations show that for randomly chosen rewiring rates $r_i$ within the two clusters, the inter-cluster 
link density reaches the same steady-state value (Fig.~\ref{5states}). A further comparison shows, that the behavior of a five-state model closely matches that of a two-state model.

Now we consider another partition, (P2), of the Jacobian, which is a partition into $s$ submatrices. Then, the corresponding generalized rowsums yield
\begin{align*}
 \mathbf{S}_i^{(\rm{P2})}&=\mathbf{\Delta}_{i}\bigl(\rho_{\underline{ij}},\rho_{1j}\bigr)+\sum_{j=1, j \neq i}^s\mathbf{\xi}_{i}(\rho_{ij})\\
=&\small{
\begin{pmatrix}
\mathbf{\tilde{D}}_{i\underline{1}}\bigl(\rho_{\underline{1j}}\bigr)& \mathbf{X}_{i\underline{1}}\bigl(\rho_{\underline{12}}\bigr) &\cdots & \mathbf{X}_{i\underline{1}}\bigl(\rho_{\underline{1s}}\bigr) \\
\mathbf{X}_{i\underline{2}}\bigl(\rho_{\underline{21}}\bigr) & \mathbf{\tilde{D}}_{i\underline{2}}\bigl(\rho_{\underline{2j}}\bigr)& \ddots & \vdots  \\
\vdots &\ddots&\ddots & \mathbf{X}_{i\underline{s-1}}(\rho_{\underline{(s-1)s}})  \\
\mathbf{X}_{i\underline{s}}\bigl(\rho_{\underline{s1}}\bigr) & \cdots & \mathbf{X}_{i\underline{s-1}}(\rho_{\underline{s(s-1)}}) &\mathbf{\tilde{D}}_{i\underline{s}}\bigl(\rho_{\underline{sj}}\bigr) 
\end{pmatrix},
}
\end{align*}
where
\begin{equation*}
\mathbf{\tilde{D}}_{i\underline i}\bigl(\rho_{\underline{ij}}\bigr)=
\mathbf{D}_{i\underline i}-\sum_{\underline{j}=\underline{1}, \underline j \neq \underline i}^{\underline{s}}\mathbf{X}_{i\underline i}(\rho_{\underline{ij}}).
\end{equation*}

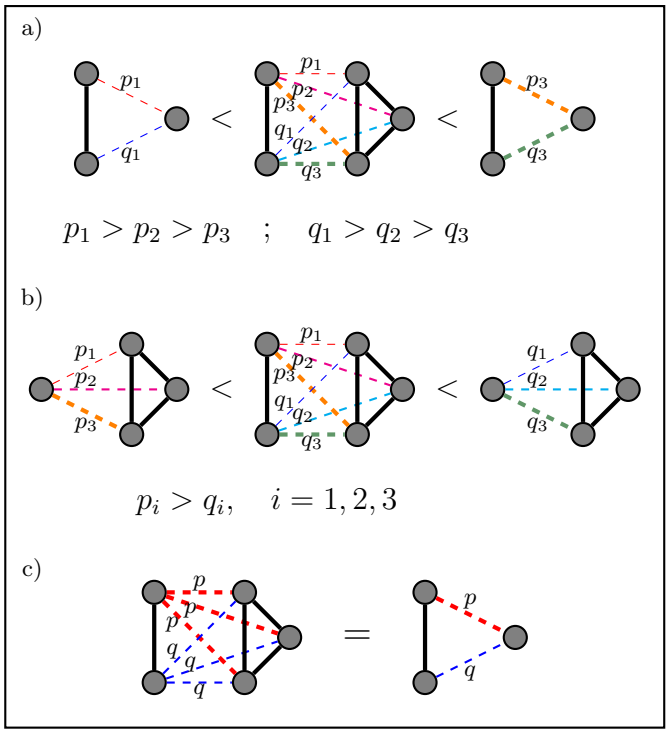
\begin{figure}
\begin{tikzpicture}[scale=0.6]

\path[use as bounding box] (-3.8-1,6.5) rectangle (8.8+1,-3.5-5);
\draw[thick] (-3.8-1,6.5) rectangle (8.8+1,-3.5-6);

\tikzstyle{grey}=[circle, thick,
                                    minimum size=0.2cm,
                                    draw=black!100,
                                    fill=black!50]
           
\draw (-3.2-1,6) node {a)};

\node[grey](A) at (0-3,0+5){};
\node[grey](B) at (0-3,-2+5){};
\node[grey](C) at (2-3,-1+5){};

\path[-](A) edge[ultra thick] (B);
\path[-](A) edge[dashed, color=red] (C);
\path[-](B) edge[dashed, color=blue] (C);

\draw (-2,4.8) node {$p_1$};
\draw (-2,3.2) node {$q_1$};

\draw[font=\large] (0,-1+5) node {$<$};

\node[grey](A) at (0+1,0+5){};
\node[grey](B) at (0+1,-2+5){};
\node[grey](C) at (2+1,0+5){};
\node[grey](D) at (2+1,-2+5){};
\node[grey](E) at (3+1,-1+5){};

\path[-](A) edge[ultra thick] (B);
\path[-](A) edge[dashed,color=red] (C);
\path[-](A) edge[ultra thick, dashed,color=orange] (D);
\path[-](A) edge[thick,dashed,color=magenta] (E);
\path[-](B) edge[dashed,color=blue] (C);
\path[-](B) edge[ultra thick, dashed,color=green] (D);
\path[-](B) edge[thick, dashed,color=cyan] (E);
\path[-](C) edge[ultra thick] (D);
\path[-](C) edge[ultra thick] (E);
\path[-](D) edge[ultra thick] (E);

\draw (3-1,5.2) node {$p_1$};
\draw (2.8-1,4.6) node {$p_2$};
\draw (2.4-1,4.3) node {$p_3$};

\draw (2.4-1,3.7) node {$q_1$};
\draw (2.8-1,3.4) node {$q_2$};
\draw (3-1,2.8) node {$q_3$};

\draw[font=\large] (5,-1+5) node {$<$}; 

\node[grey](A) at (0+8-2,0+5){};
\node[grey](B) at (0+8-2,-2+5){};
\node[grey](C) at (2+8-2,-1+5){};

\path[-](A) edge[ultra thick] (B);
\path[-](A) edge[ultra thick, dashed, color=orange] (C);
\path[-](B) edge[ultra thick, dashed, color=green] (C);

\draw (9-2,4.8) node {$p_3$};
\draw (9-2,3.2) node {$q_3$};

\draw[font=\large] (1,1.5) node {$p_1>p_2>p_3\quad; \quad q_1>q_2>q_3$};

\draw (-3.2-1,6-6) node {b)};

\node[grey](A) at (0-3-1,0+5-6-1){};
\node[grey](C) at (2+1-4-1,0+5-6){};
\node[grey](D) at (2+1-4-1,-2+5-6){};
\node[grey](E) at (3+1-4-1,-1+5-6){};

\path[-](A) edge[dashed, color =red] (C);
\path[-](A) edge[dashed, color=orange, ultra thick] (D);
\path[-](A) edge[dashed, color=magenta, thick] (E);
\path[-](C) edge[ultra thick] (D);
\path[-](C) edge[ultra thick] (E);
\path[-](D) edge[ultra thick] (E);

\draw (-2-1,4.8-6) node {$p_1$};
\draw (-2-1,4.2-6) node {$p_2$};
\draw (-2-1,3.2-6) node {$p_3$};

\draw[font=\large] (0,-1+5-6) node {$<$};

\node[grey](A) at (0+1,0+5-6){};
\node[grey](B) at (0+1,-2+5-6){};
\node[grey](C) at (2+1,0+5-6){};
\node[grey](D) at (2+1,-2+5-6){};
\node[grey](E) at (3+1,-1+5-6){};

\path[-](A) edge[ultra thick] (B);
\path[-](A) edge[dashed,color=red] (C);
\path[-](A) edge[ultra thick, dashed,color=orange] (D);
\path[-](A) edge[thick,dashed,color=magenta] (E);
\path[-](B) edge[dashed,color=blue] (C);
\path[-](B) edge[ultra thick, dashed,color=green] (D);
\path[-](B) edge[thick, dashed,color=cyan] (E);
\path[-](C) edge[ultra thick] (D);
\path[-](C) edge[ultra thick] (E);
\path[-](D) edge[ultra thick] (E);

\draw (3-1,5.2-6) node {$p_1$};
\draw (2.8-1,4.6-6) node {$p_2$};
\draw (2.4-1,4.3-6) node {$p_3$};

\draw (2.4-1,3.7-6) node {$q_1$};
\draw (2.8-1,3.4-6) node {$q_2$};
\draw (3-1,2.8-6) node {$q_3$};

\draw[font=\large] (5,-1+5-6) node {$<$}; 

\node[grey](A) at (0-3+8+1,0+5-6-1){};
\node[grey](C) at (3+1+4,0+5-6){};
\node[grey](D) at (3+1+4,-2+5-6){};
\node[grey](E) at (4+1+4,-1+5-6){};

\path[-](A) edge[dashed, color=blue] (C);
\path[-](A) edge[dashed, color=green, ultra thick] (D);
\path[-](A) edge[dashed, color=cyan, thick] (E);
\path[-](C) edge[ultra thick] (D);
\path[-](C) edge[ultra thick] (E);
\path[-](D) edge[ultra thick] (E);

\draw (9-2,4.8-6) node {$q_1$};
\draw (9-2,4.2-6) node {$q_2$};
\draw (9-2,3.2-6) node {$q_3$};

\draw[font=\large] (1,1.5-6) node {$p_i>q_i, \quad i=1,2,3$};

\draw (-3.2-1,0-6) node {c)};

\node[grey](A) at (0-1.5,0-.5-6){};
\node[grey](B) at (0-1.5,-2-.5-6){};
\node[grey](C) at (2-1.5,0-.5-6){};
\node[grey](D) at (2-1.5,-2-.5-6){};
\node[grey](E) at (3-1.5,-1-.5-6){};

\path[-](A) edge[ultra thick] (B);
\path[-](A) edge[ultra thick, dashed, color=red] (C);
\path[-](A) edge[ultra thick, dashed, color=red] (D);
\path[-](A) edge[ultra thick, dashed, color=red] (E);
\path[-](B) edge[thick, dashed, color=blue] (C);
\path[-](B) edge[thick, dashed, color=blue] (D);
\path[-](B) edge[thick, dashed, color=blue] (E);
\path[-](C) edge[ultra thick] (D);
\path[-](C) edge[ultra thick] (E);
\path[-](E) edge[ultra thick] (D);

\draw (3-3.5,5.2-5-.5-6) node {$p$};
\draw (2.8-3.5,4.6-5-.5-6) node {$p$};
\draw (2.4-3.5,4.3-5-.5-6) node {$p$};

\draw (2.4-3.5,3.7-5-.5-6) node {$q$};
\draw (2.8-3.5,3.4-5-.5-6) node {$q$};
\draw (3-3.5,2.8-5-.5-6) node {$q$};

\draw[font=\Large] (5-2,-1-.5-6) node {$=$};

\node[grey](A) at (0+6-1.5,0-.5-6){};
\node[grey](B) at (0+6-1.5,-2-.5-6){};
\node[grey](C) at (2+6-1.5,-1-.5-6){};

\path[-](A) edge[ultra thick,] (B);
\path[-](A) edge[ultra thick, dashed, color=red] (C);
\path[-](B) edge[thick, dashed, color=blue] (C);

\draw (7-1.5,-0.2-.5-6) node {$p$};
\draw (7-1.5,-1.8-.5-6) node {$q$};

\end{tikzpicture}
\caption{Schematic representation of the generalized Frobenius inequalities using the example of a five-state system and a specific partial fragmentation into a 2-state- and a 3-state-cluster.
The inequalities in a) and b) represent two different sets of bounding systems obtained from a partition of type (P2), as described in the text.
In a) the bounding systems are three-state systems, obtained by reducing the 3-cluster to a single state. The remaining inter-cluster rewiring rates are the respective maximum (minimum) values
of the rewiring rates between each state in the 2-cluster and all states in the 3-cluster.
In b) the bounding systems are four-state systems, obtained by reducing the 2-cluster to a single state. The remaining inter-cluster rewiring rates are the maximum (minimum) values
of the rewiring rates between each state in the 3-cluster and all states in the 2-cluster.
In c) a special case is shown where the upper and lower three-state bounding systems coincide. In this case, the leading eigenvalue of the five-state system equals the leading eigenvalue of the associated three-state system and the considered fragmentation of the 
original system is fully captured by the lower-dimensional one.}
\label{equival}
\end{figure}

First, we observe that every matrix $\mathbf{S}_{i}^{\rm{(P2)}}$ corresponds to a partial fragmentation in a system of $\overline{s}+1$ states.
More precisely, the set $\{\mathbf{S}_{i}^{\rm{(P2)}}\}$ describes a collection of $s$ single-state-fragmentations where for every $i$ a single state $g_i$ is taken separately from the $s$-cluster. This single state (i.e. now $s=1$) then forms the first component
of the partial fragmentation, while the second component is given by the whole $\underline{s}$-cluster.

Now, building the element-wise extrema of $\{\mathbf{S}_{i}^{\rm{(P2)}}\}$ means to compare all the single-state-fragmentations by comparing every matrix-entry of the corresponding generalized rowsums.
Taking $\min_i\mathbf{S}_{i}^{\rm{(P2)}}$ ($\max_i\mathbf{S}_{i}^{\rm{(P2)}}$) yields therefore in evere matrix-entry the minimum (maximum) value, i.e.{} that one which comprises the maximal (minimal) rewiring rate.
The resulting matrix corresponds to a partial fragmentation with respect to a single state where the inter-cluster rewiring rates are chosen extremal according to the described comparison.
We will refer to such a system as \emph{bounding system} (see Fig.~\ref{equival} for exemplary bounding systems).

For a partition of type (P2) the leading eigenvalue of the general Jacobian satisfies
\begin{equation}
\label{ineqjac2}
 \lambda\bigl(\min_i\mathbf{S}_{i}^{\rm{(P2)}}\bigr) \le \lambda\bigl(\mathbf{J})\le \lambda(\max_i\mathbf{S}_{i}^{\rm{(P2)}}\bigr).
\end{equation}
The lower bound corresponds to the fragmentation of a system where the largest inter-cluster rewiring rate of each state in the second component is connected to a single state.
The upper bound  corresponds to the fragmentation of a system where the smallest inter-cluster rewiring rate of each state in the second component is connected to a single state.

As in the Jacobian in \eqref{genjac} the matrices $\mathbf{X}$ and $\mathbf{X^\prime}$ can be interchanged, one can consider a corresponding partition where the second component is reduced to a single state, i.e.{} $\underline{s}=1$ and the
first component remains as a whole. This leads to a different set of bounding systems, as illustrated in Fig.~\ref{equival} a) and b).

From (\ref{ineqjac2}) we can draw the following conclusions:
\begin{itemize}
 \item If the lower bounding system does not fragment ($\lambda\bigl(\min_i\mathbf{S}_{i}^{\rm{(P2)}}\bigr)>0$) the original system does not fragment.
\item If the upper bounding system fragments ($\lambda(\max_i\mathbf{S}_{i}^{\rm{(P2)}}\bigr)<0$) the original system fragments.
\item If upper and lower bounding systems are the same ($\min_i \mathbf{S}_{i}^{\rm{(P2)}}=\max_i \mathbf{S}_{i}^{\rm{(P2)}}$) the fragmentation of the original system is exactly captured by the bounding system, i.e.
the full Jacobian $\mathbf{J}$ reduces to the Jacobian of the bounding system. 
\end{itemize}
The latter case is realized if every state in one component is connected via equal rewiring rates to every state in the other component (see Fig.~\ref{equival} c)). For state-network topologies which display this property 
the dimension of the Jacobian reduces significantly and thus the fragmentation condition becomes much more tractable.

In summary, the results from the second partition show that for the leading eigenvalue of a Jacobian, corresponding to a partial fragmentation into two active clusters,
upper and lower bounds can be given, which correspond to single-state-fragmentations in (properly constructed) lower-dimensional systems. 
In particular, the leading eigenvalue of the full Jacobian can be exactly calculated as the leading eigenvalue of a lower-dimensional Jacobian if special state-network topologies are given.
Otherwise, when such a reduction is not possible, the bounding systems provide necessary conditions for a partial fragmentation to occur. So, calculating fragmentation conditions for the much simpler bounding systems in some cases
suffices to predict the ocurrence or absence of fragmentations in the full system.

\section{CONCLUSION}
In the present paper we extended recent work on the adaptive two-state voter models to a family of multi-state models.
For the three-state model our analysis revealed a phase diagram in which three distinct types of long-term behavior are observed. 
Depending on the parameters the system either approaches a consensus state, a partially fragmented state ultimately leading to two surviving 
opinions or a fully fragmented state in which all three opinions survive.  

In a general scenario with an arbitrary number of states making precise predictions is more difficult. 
In particular, the computation of fragmentation points generally requires the estimation of active link densities inside the clusters between which the 
fragmentation occurs. By exploiting the specific structure of transition rates in the system, one can nevertheless gain analytical insights 
into the fragmentation dynamics. For example we identified a class of special cases in which adaptive multi-state voter models exactly recover 
the behavior of the adaptive two-state voter model.     

While the ultimate goal of understanding opinion formation in the human population is still far away, the present progress shows 
that analytical understanding can be pushed to more complex models. Recent studies have shown that already today variants of the voter model 
can be tested in experiments with swarming animals. An important goal for the future is to continue the refinement of models and analysis techniques in order to describe real-world situations. We hope that the approach presented here will make a contribution to this ongoing process.

\section{APPENDIX}
We use a moment closure approach \cite{keeling1999} for the calculation of the maximal active link density in an active cluster of $s$ states in a $G$-state system. The evolution equation for the number of active links of type $xy$ is given by
\begin{equation*}
 \dot{[xy]}=-[xy]+\frac{1}{2}\bigl( \sum_{z\neq x} [xzy]+\sum_{z\neq y} [xzy] -\sum_{z\neq y} [xyz] -\sum_{z\neq x} [zxy]\bigr),
\end{equation*}
assuming $p_{ij}=0$ for all rewiring rates within the active cluster. Then, using the pair-approximation, we get for the steady-state,
\begin{align}
 [xy]=\frac{1}{2}&\Bigl(\sum_{z\neq x,y} 2\frac{[xz][zy]}{[z]}+2\frac{[xy][yy]}{[y]}+2\frac{[xx][xy]}{[x]}\Bigr.\nonumber\\
\Bigl.&-\sum_{z\neq y}\frac{[xy][yz]}{[y]}-\sum_{z\neq x} \frac{[zx][xy]}{[x]}\Bigr).
\label{xydens}
\end{align}
For equal distribution of states we can write $[x]=n/s\, \forall x, [xy]=\zeta\, \forall (x,y)$ and $[xx]=\eta\, \forall x$, where $n$ denotes the number of nodes in the active component.
The total number of links in the active component $l$ is then given by
\begin{equation*}
 l=s\eta+\frac{s(s-1)}{2}\zeta
\end{equation*}
and (\ref{xydens}) yields
\begin{eqnarray*}
 \zeta&=&\frac{\zeta^2}{n}s(s-2)+2\frac{\zeta\eta}{n}s-\frac{\zeta^2}{n}s(s-1)\\
&& \frac{\zeta^2}{n}s(s-2)+2\frac{\zeta l}{n}-2\frac{\zeta^2}{n}s(s-1).
\end{eqnarray*}
Using $kn=2l$, we get for the maximum active link density in a cluster of $s$ states:
\begin{equation*}
 \rho_{\max}=\frac{s(s-1)\zeta}{2l}=\frac{(s-1)(k-1)}{sk},
\end{equation*}
which gives $\rho_{\max}=(k-1)/(2k)$ for $s=2$, as provided in the text.


\end{document}